\documentstyle[prc,aps,preprint]{revtex}
\title{The Determination of Nuclear Level Densities from Experimental
Information}
\date{\today}
\draft
\author{B. J. Cole}
\address{Department of Physics, University of the Witwatersrand,
     WITS 2050, South Africa}
\author{N. J. Davidson}
\address{Department of Mathematics, University of Manchester
Institute of
Science and Technology (UMIST), P O Box 88, Manchester M60 1QD,
England}
\author{H. G. Miller}
\address{Department of Physics,
        University of Pretoria, Pretoria 0002, South Africa}

\begin{document}

\maketitle
\begin{abstract}
A novel Information Theory based method for determining the density
of states
from prior information is presented. The energy dependence of the
density of
states is determined from the observed number of states per energy
interval
and model calculations suggest that the method is sufficiently
reliable to
calculate the thermal properties of nuclei over a reasonable
temperature range.

\end{abstract}


The simplest expression for the nuclear level density has been
obtained in
the Fermi-gas model by Bethe~\cite{B36,B37,B38} and later modified by
Bloch~%
\cite{B54,LL54,E60}. There are, however, a number of shortcomings in
this
approach. For example, the lack of coupling to the collective part of
the
nuclear spectrum leads to an energy-independent level density
parameter.
Recently there has been considerable theoretical activity in the
determination of the nuclear many-body density of states, taking into
account shell, pairing, and deformation effects~\cite{GC65,W80},
finite size
effects~\cite{YM92}, and thermal and quantal
effects~\cite{PBB90,PBB91}
as well
as improvements in the determination of the spin cut-off
factors~\cite{G80}.
Furthermore, the multiple inverse Laplace transform used to
determine the nuclear density of states from the Grand-Canonical
partition function of a Fermi gas appears to lead to
certain inconsistencies in the folding of nuclear level
densities~\cite{GH93}. In spite of these deficiencies Bloch's formula
is widely used,
particularly as a means of parameterizing the experimentally
determined
nuclear level densities~\cite{EFS61}. Generally the level density
parameter
is taken to be constant. Should it be energy dependent, an {\em a
priori}
form for
the energy dependence must be assumed.

In the present work we will demonstrate a simple means of extracting
the nuclear level density from the experimental information, based on
Information Theory~\cite{MJ86}. We wish to determine the distribution of states
or level density, $\rho (E),$ for a nucleus such
that
\begin{equation}
\label{rho}\int_0^\infty \rho (E)\Theta (E_i-E)\Theta (E-E_{i+1})
dE=N_i%
\hspace{5mm}i=1,\ldots ,N
\end{equation}
where $N_i,$ the number of states in the energy interval $E_{i+1}-
E_i$, is
assumed known, and $\Theta \left( x\right) =1$ if $x\geq 0$ and is
zero
otherwise. As a zeroth-order guess we assume that the distribution of
states
or level density is given by the Bloch formula
\begin{equation}
\label{rhozero}\rho _0(E)=\frac 1{\sqrt{48}E}\exp \left[
2\sqrt{a_0E}\right]
\end{equation}
where $a_0$ is the level-density parameter, often expressed in terms
of the
nuclear mass number $A$ by
\begin{equation}
\label{mass}a_0=\frac A{k_0},
\end{equation}
with, for example, $k_0$ = 8 MeV. The actual distribution $\rho (E)$
may now
be determined by minimizing the information entropy of $\rho (E)$
relative
to $\rho _0(E)$
\begin{equation}
S(\rho ,\rho _0)=\int\limits_0^\infty \rho (E)\ln \frac{\rho
(E)}{\rho _0(E)}%
dE
\end{equation}
subject to the experimental information given in eq.~(\ref{rho}).
This
yields
\begin{equation}
\label{rhocalc}\rho (E)=\rho _0(E)\exp \left[ \sum_{i=1}^N
\lambda_i\Theta
(E_i-E)\Theta (E-E_{i+1})\right]
\end{equation}
where the Lagrange multipliers $\lambda _i$ for the intervals
$E_i\leq E\leq
E_{i+1}$ are determined from the experimental information given in
eq.~(\ref
{rho}).

Now suppose we have determined the set of $\lambda _i$ and we put
\begin{equation}
\lambda (\bar E_i)=\lambda _i
\end{equation}
where $\bar E_i$ is the midpoint of the energy interval $E_i\leq
E\leq
E_{i+1}$. If the $\lambda (\bar E_i)$ vary smoothly with energy, a
least-squares fit to this quantity should enable us to extrapolate
the $\rho
(E)$ beyond the region in which the experimental information is
known. If
this is the case the density of states may now given by
\begin{equation}
\rho (E)=\rho _0(E)e^{-\lambda (E)},
\end{equation}
or, from eqs (\ref{rhozero}) and (\ref{mass}),
\begin{equation}
\label{rhocalc2}\rho \left( E\right) =\frac 1{\sqrt{48}E}\exp \left[
2\sqrt{
\frac{AE}{k_0}}-\lambda \left( E\right) \right] .
\end{equation}
Note  here that the energy dependence of the density of states is
determined
from the prior experimental data.

Thermal properties of nuclei are calculated from the nuclear
partition
function $Z\left( T\right) $ for fixed nuclear mass number $A$, where
$T$ is
the nuclear temperature. The contribution to $Z\left( T\right) $ of
the
continuous part of the nuclear spectrum is
\begin{equation}
\label{part}Z\left( T\right) =\int_{E_{\min }}^\infty \rho \left(
E\right)
e^{-E/T}dE
\end{equation}
where $E_{\min }$ is some suitable lower limit.

To test the proposed method we have performed the following model
calculations. Rather than use actual experimental information for the
number
of states $N_i$ in eq. (\ref{rho}), we have generated the $N_i$ by
assuming
that for energy $E\geq 1$ MeV the experimental distribution of states
is
given by
\begin{equation}
\label{rhoinp}\rho \left( E\right) =\frac 1{\sqrt{48}E}\exp \left[
2\sqrt{
\frac{AE}{k\left( E\right) }}\right]
\end{equation}
where $k=k(E)$ to allow the energy dependence of the density of
states to
deviate from the Bloch form. The size of the energy intervals
$E_{i+1}-E_i$
in eq. $\left( \ref{rho}\right) $ was fixed at 0.4 MeV, with ten such
intervals covering the energy range 1--5 MeV. For a particular choice
$k(E)$
the Lagrange multipliers $\lambda _i$ were calculated from eqs
(\ref{rho}), (%
\ref{rhozero}), (\ref{mass}) and (\ref{rhocalc}), with $A$
arbitrarily set
to 56 in eq. (\ref{mass}). A series of least-squares polynomial fits
of low
order was then performed for the modified Lagrange multipliers
\begin{equation}
\lambda ^{\prime }\left( \bar E_i\right) =\frac{\lambda \left( \bar
E_i\right) }{2\sqrt{AE}}.
\end{equation}
The many-body density of states and nuclear partition function were
then
calculated from eqs $\left( \ref{rhocalc2}\right) $ and $\left(
\ref{part}%
\right) ,$ respectively.

A sample fit to $\lambda ^{\prime }\left( \bar E_i\right) $ is shown
in
figure 1 for the choice
\begin{equation}
\label{choice1}k(E)=8
\end{equation}
in eq. $\left( \ref{rhoinp}\right) $, corresponding to the Bloch
formula (%
\ref{rhozero}) with $k_0$ = 8; a perfect fit  should produce $\lambda
^{\prime }$ = 0 identically.  Since we take the number of states to
be
an integer, whereas the integral in (\ref{rho}) with the density of
states
given by (\ref{rhoinp}) produces a real number, we do not expect to
find
all the Lagrange multipliers to be zero. The error bars included in
the figure
were generated by arbitrarily changing the number of states $N_i$ to
$N_i \pm \sqrt{%
N_i/2}.$ They serve both to indicate the sensitivity of the
calculation to
uncertainties in the input, and also to provide a scale for $\lambda
^{\prime }$. The fit was also weighted according to these error bars.
As
figure 1 indicates, the quality of the fit increases as the order of
the
polynomial is increased, but $\lambda ^{\prime }=4\times 10^{-5}$
(constant)
is already satisfactory.

The corresponding many-body density of states was calculated from
eq. (\ref
{rhocalc2}) with $E$ extrapolated beyond the region of fit 1--5 MeV
to 20
MeV; the density relative to the input density eq. $\left(
\ref{rhoinp}%
\right) $ is shown in figure 2. It is seen that $\lambda ^{\prime }$
=
constant provides an almost perfect fit for all energies.  Polynomial
fits
of higher order, whilst giving a slightly better fit up to $E$ = 5
MeV,
cannot be extrapolated much beyond this energy.

The nuclear partition function was computed using eq. $\left(
\ref{part}%
\right) $, with $E_{\min }=1$ MeV. Due to the rapid increase in the
calculated density of states for higher order polynomial fits to the modified
Lagrange multipliers, the infinite upper limit of the
 energy
integration was replaced by $E_{\max }$, determined as follows.
First,
$Z(T)$
for a given temperature was calculated using the input density of
states eq.
$\left( \ref{rhoinp}\right) $, with the upper limit  gradually
increased
until the integral converged to a constant value.  The final value
for the
upper limit, $E_{\max }$, which varied between 30 MeV and 150 MeV
depending
on $T$, was then used in the calculation for that temperature with
the
fitted density of states eq. $\left( \ref{rhocalc2}\right) $. The
results
are plotted in figure 3, which shows $Z(T)$ calculated with eq.
$\left( \ref
{rhocalc2}\right) $ relative to the partition function calculated
with the
input density of states. The fit $\lambda ^{\prime }$ = constant
produces an
almost perfect result up to $T$ = 2 MeV, whereas the higher order
fits are
far from satisfactory except at the lowest temperatures.

These calculations were repeated with other choices for $k(E)$ in eq.
$%
\left( \ref{rhoinp}\right) $. For example, the use of
\begin{equation}
\label{choice2}k(E)=6+E/2
\end{equation}
is illustrated in figures 4--6. This choice allows a reasonable
energy
dependence for the level density parameter, permitting the density of
states
to deviate sufficiently from the Bloch form to provide a realistic
test of
the method. (It should be noted that in practice the energy
dependence of $%
k(E)$ is constrained; too strong a dependence causes the density of
states
to increase rapidly at higher energies or to decrease suddenly to
zero.) As
seen from figure 4, the lowest order polynomial which fits $\lambda
^{\prime
}\left( \bar E_i\right) $ within the error bars is linear in energy.
Figure
5  shows that this same fit provides a satisfactory description of
the
density of states well beyond the energy region in which the fit was
produced. The quality of the calculated partition function, shown if
figure
6, is excellent.

Finally, we have checked the calculations for sensitivity to various
details
in the method. For example we have performed calculations in which:

\begin{itemize}
\item  the energy $\bar E_i$ for the energy interval $E_{i+1}-E_i$
was
computed as the mean energy weighted according to the density of
states,
rather than simply as the midpoint energy;

\item  the fit to $\lambda ^{\prime }\left( \bar E_i\right) $ was
{\em not}
weighted according to the uncertainty in $\lambda ^{\prime }$;

\item  the fit was to $\lambda \left( \bar E_i\right) $ rather than
to
$\lambda ^{\prime }\left( \bar E_i\right) $;

\item  the energy range 1--5 MeV was spanned by 5 intervals of width
0.8
MeV, or 16 intervals of width 0.25 MeV, rather than 10 intervals of
width
0.4 MeV.
\end{itemize}
In most cases the results were not changed significantly. There is a
slight
preference for fewer, wider energy intervals, and for the fit to
$\lambda
^{\prime }$ to be weighted. A fit to $\lambda ^{\prime }$, rather
than
$\lambda $, is definitely preferred, at least for those functions
$k(E)$
actually used.

A simple Information theory based method for determining the density
of
states from prior experimental measurements has been presented. A
good
approximation of the density of states  is obtained even at higher
energies
and any energy dependence is determined from the prior experimental
data.   However, the extrapolated density of states will not
necessarily be
very accurate if at higher energies processes occur which are not
reflected
in the lower energy prior information. With this proviso, the model
calculations show that the partition function can be accurately
calculated
over a reasonable temperature range which suggest that the thermal
properties
of nuclei can be adequately described by this method.

\acknowledgments
BJC and HGM acknowledge the financial support of the Foundation for
Research Development, Pretoria.

\begin{figure}
\caption{Modified Lagrange multipliers for the case $k(E)$ = 8.
The circles indicate values computed with eq. (8 ),
the solid line represents the fit with a polynomial of order zero,
the dashed line the fit of order one
and the dash-dot line the fit of order two. The calculation of the
error bars is explained in the text.}
\end{figure}

\begin{figure}
\caption{Calculated density of states, relative
to the exact density of states, for the case $k(E)$ = 8.
The solid line is computed with $\lambda$ fitted with a polynomial of
order zero,
the dashed line with the fit of order one
and the dash-dot line with the fit of order two.}
\end{figure}

\begin{figure}
\caption{Calculated partition function, relative
to the exact partition function, for the case $k(E)$ = 8.
The solid line is computed with $\lambda$ fitted with a polynomial of
order zero,
the dashed line with the fit of order one
and the dash-dot line with the fit of order two.}
\end{figure}

\begin{figure}
\caption{Modified Lagrange multipliers for the case $k(E)=6+E/2$.
The circles indicate values computed with eq. (8),
the solid line represents the fit with a polynomial of order zero,
the dashed line the fit of order one
and the dash-dot line the fit of order two. The calculation of the
error bars is explained in the text.}
\end{figure}

\begin{figure}
\caption{Calculated density of states, relative
to the exact density of states, for the case $k(E)=6+E/2$.
The solid line is computed with $\lambda$ fitted with a polynomial of
order zero,
the dashed line with the fit of order one
and the dash-dot line with the fit of order two.}
\end{figure}

\begin{figure}
\caption{Calculated partition function, relative
to the exact partition function, for the case $k(E)=6+E/2$.
The solid line is computed with $\lambda$ fitted with a polynomial of
order zero and
the dashed line with the fit of order one.}
\end{figure}

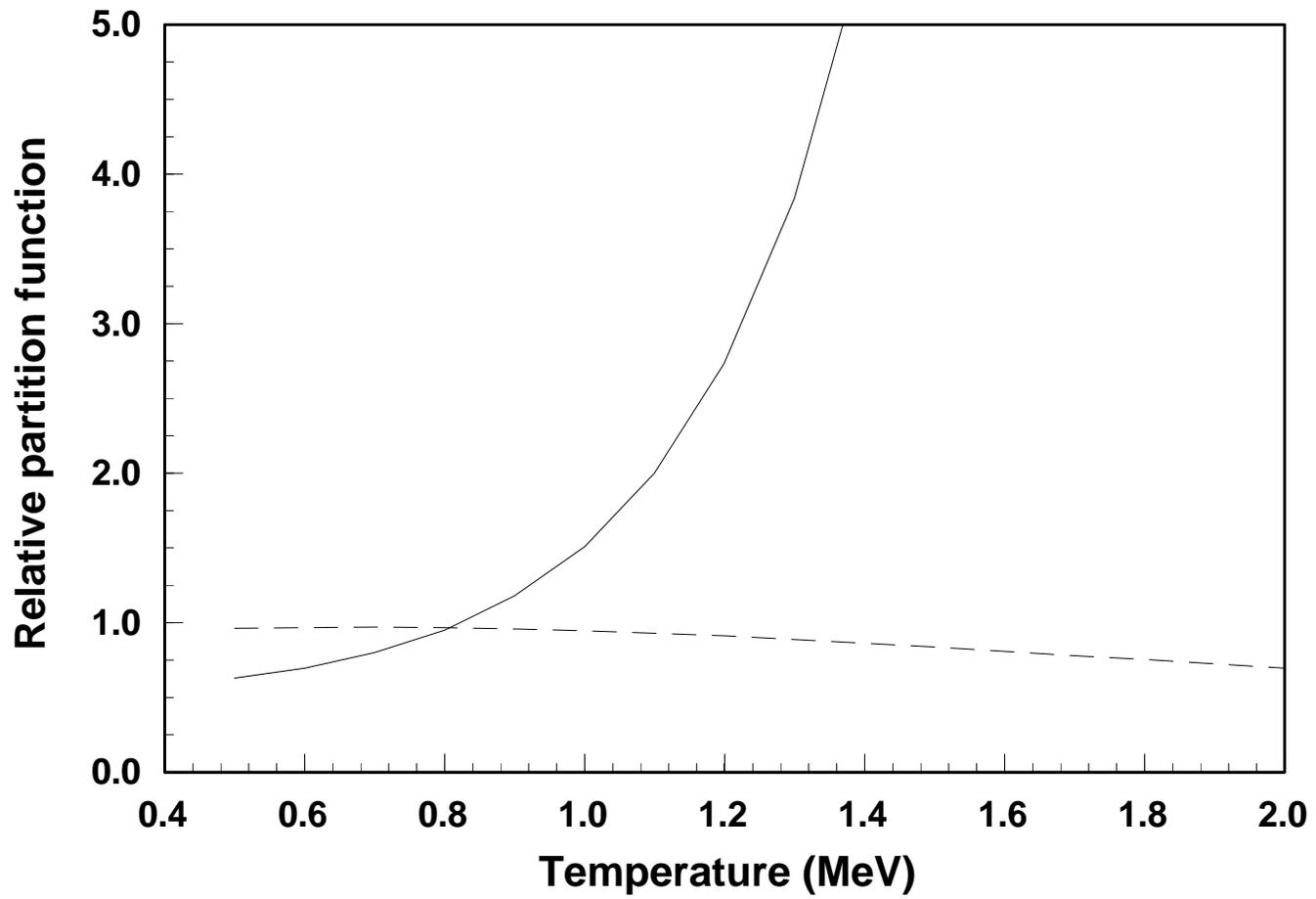
\end{document}